# Understanding Global Galactic Star Formation

Paul Scowen

School of Earth & Space Exploration
Arizona State University
PO Box 871404, Tempe, AZ 85287-1404

(480) 965-0938

paul.scowen@asu.edu

Rolf Jansen (Arizona State University)

Matthew Beasley (University of Colorado – Boulder)

Daniela Calzetti (University of Massachusetts)

Steven Desch (Arizona State University)

John Gallagher (University of Wisconsin – Madison)

Mark McCaughrean (University of Exeter)

Robert O'Connell (University of Virginia)

Sally Oey (University of Michigan)

Deborah Padgett (IPAC / Caltech)

Aki Roberge (NASA – GSFC)

Nathan Smith (University of California – Berkeley)

Science White Paper for the Astro2010 Decadal Survey
Panel(s): Planetary Systems and Star Formation; Stars and Stellar Evolution

**Abstract**

We propose to the community a comprehensive UV/optical/NIR imaging survey of Galactic star formation regions to probe all aspects of the star formation process. The primary goal of such a study is to understand the evolution of circumstellar protoplanetary disks and other detailed aspects of star formation in a wide variety of different environments. This requires a comprehensive emission-line survey of nearby star-forming regions in the Milky Way, where a high spatial resolution telescope+camera will be capable of resolving circumstellar material and shock structures. In addition to resolving circumstellar disks themselves, such observations will study shocks in the jets and outflows from young stars, which are probes of accretion in the youngest protoplanetary disks still embedded in their surrounding molecular clouds. These data will allow the measurement of proper motions for a large sample of stars and jets/shocks in massive star-forming regions for the first time, opening a new window to study the dynamics of these environments. It will require better than 30 mas resolution and a stable PSF to conduct precision astrometry and photometry of stars and nebulae. Such data will allow production of precise color-color and color magnitude diagrams for millions of young stars to study their evolutionary states. One can also determine stellar rotation, multiplicity, and clustering statistics as functions of environment and location in the Galaxy. For the first time we can systematically map the detailed excitation structure of HII regions, stellar winds, supernova remnants, and supershells/superbubbles. This survey will provide the basic data required to understand star formation as a fundamental astrophysical process that controls the evolution of the baryonic contents of the Universe.

**Introduction & Scientific Context**

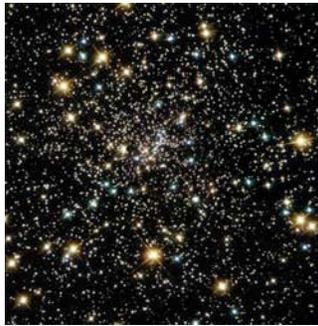

Stars are the fundamental building blocks of the Universe and influence its evolution on all scales, from the cosmological to the planetary. The formation of stars locks away baryons for a Hubble time, they produce the energy that establishes the state of matter in the interstellar medium (ISM), they control the fate of self-gravitating masses, and they produce the light that renders distant galaxies visible. It is because of stars that elements heavier than helium are created. Without stars, there would be no planets, no carbon, and no free energy to drive the evolution of life. Star formation is *the* fundamental process underpinning the evolution of the Universe and life within it. Progress toward understanding the cosmic history of normal matter, the formation and evolution of galaxies, the birth and fate of planetary systems, and our own origins requires a comprehensive understanding of star formation *as a large-scale, coherent, systematic process.*

There has been remarkable progress in our understanding of star formation during recent decades. Molecular clouds form from the ISM. Their densest cores suffer gravitational collapse to form protostars which are $10^7$ times smaller and 21 orders of magnitude denser. Spin and pressure gradients channel accretion from the envelope onto a spinning disk. Magnetic fields grow, extract angular momentum, and drive accretion from the disk onto the star. Dynamo-generated stellar magnetic fields regulate stellar rotation, accrete matter onto the star at high latitudes, and expel supersonic jets and bipolar outflows. Particles in the disk grow, sediment, and eventually form planets around the young star.

Observations, theory and numerical simulations have led to major paradigm shifts in this simple description of star and planet formation. First, the birth of isolated stars from a quiescent dark cloud is rare. Observations have shown that most stars form in turbulent giant molecular clouds with supersonic motions having Mach numbers of 10 to 100. Second, most stars form in dense clusters in close proximity to tens, hundreds, or even thousands of other stars. Some siblings are massive stars with powerful stellar winds, intense UV radiation fields, and violent and explosive deaths that dramatically affect the surrounding ISM. The vast majority of normal stars, probably including our own Sun, formed in such OB associations. Feedback of light, energy, and matter drives and regulates cloud formation, gravitational collapse, and the properties of the individual stars, multiple systems, and the clusters that form. These stochastic turbulent processes appear to be fundamental to understanding the origin and distribution of stellar masses and other stellar properties.

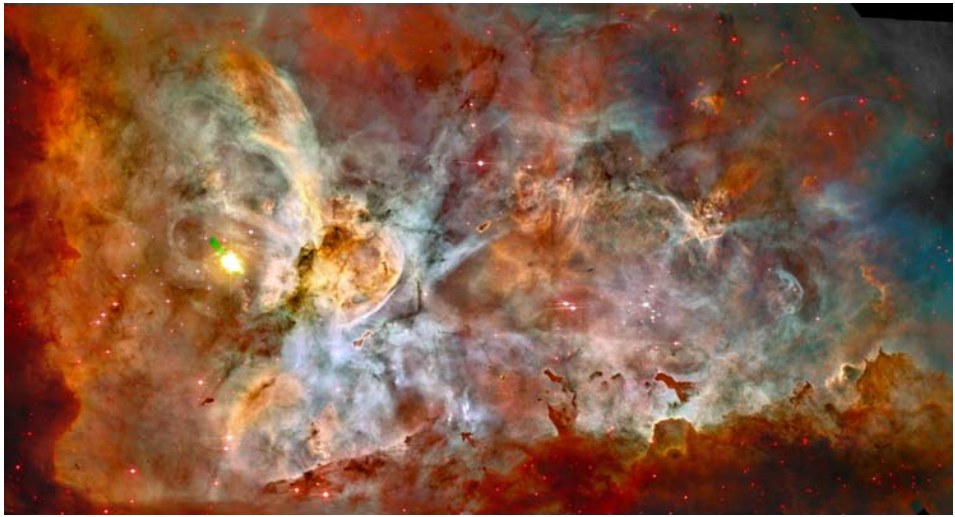

**Figure 1** – the HST mosaic of the Carina Nebula (Smith et al 2008). This is the kind of dataset that, when replicated across all massive star forming regions within 2-2.5 kpc of the Sun, will yield a dataset capable of unlocking the secrets of star formation as a global process

**Compelling Science Themes Based on Recent Advances**

We believe that to understand and address star formation as a global *system*, we need to design and engage in a systematic program of imaging that covers a large number and variety of Galactic star forming regions. To understand star birth in the early Universe, to understand galaxy formation and evolution, to understand the origin of the stellar mass spectrum, to understand the formation of planets, and to understand feedback, we must treat star birth as an integrated systemic process. We must observe star forming complexes in their entirety: we must trace the interactions between gas and stars, between stars and stars, and between disks and their environments. To make progress, we must spatially resolve disks, multiple stars, and star clusters. We must measure stellar motions, and perform relative photometry with sufficient precision to age-date young stars. All these top-level goals make specific requirements of any instrumentation designed to execute this program – requirements that we will detail in subsequent sections. At the heart of this program is the goal of providing critical advances in our knowledge of star and planet birth.

The goals of our Galactic star forming imaging program are to make major advances in the following topics:

**Young stellar objects** (YSOs): Masses, mass-spectra, rotation rates, variability, ages, multiplicity, clustering statistics, motions, brown dwarfs, free-floating proto-planets.

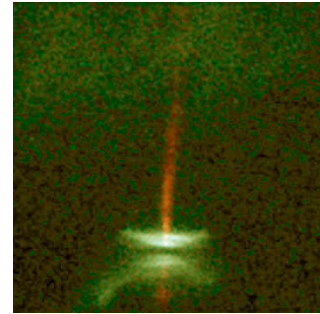

We need to be able to trace individual star, multiple star, and cluster properties to assay the range of star formation products and the manner in which they are assembled – a goal that requires the combination of a wide field of view, high angular resolution, and photometric and astrometric stability potentially enabling sub-milli-arcsecond relative astrometry, and milli-magnitude relative photometry. Measurement of the orbital motions of stars is necessary to gain insight into the dynamics of stars once they are produced, how cluster dispersion varies, and the possible detection of high velocity stars, as well as mapping large-scale nebular motions. Such measurements all require kilometer-per-second proper motion sensitivity for both stars and compact nebulae. Measurement of the stellar rotation rates for most stars is necessary to understand the resulting dynamics of each star formation episode and is achieved by recording star-spot modulation using precise relative photometry. Of particular interest is the search for transiting proto-planets in a subset of edge-on disks, and with a large-area imaging survey we will capture extremely rare types of events such as proto-planet collisions in 1 to 100 Myr old debris disks in associations. Precise cluster and association ages will be determined by fitting of HR diagram turn-on and turn-off loci requiring accurate relative photometry. Extending this same photometry to binaries will enable the best calibrations of pre-main sequence evolutionary tracks. Addressing the questions of clustering, young cluster evolution, and cluster dissipation will require stellar positions and motions to be probed. With such datasets we will identify many young brown-dwarfs and free-floating protoplanets.

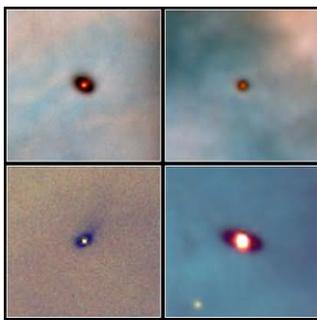

**Disks**: Sizes, masses, structure, mass-loss rates, photo-evaporation, density distributions, survival times.

A primary goal is to identify thousands of protoplanetary disks seen in silhouette, and embedded within evaporating proplyd envelopes in dozens of nearby HII regions, out to a distance of about 2 kpc. The widefield survey images taken toward regions such as Orion or Carina will extend the surveyed areas by one to two orders of magnitude over the most ambitious HST surveys undertaken so far. It will be possible to sample disks with ages ranging from 0.1 Myr to over 100 Myr when a variety of selected lines of sight are observed toward the Perseus, Orion, and Carina regions. It will be possible to look for spiral structure, gaps, and other evidence for disk perturbations from both internal and external influences. The nearest disks are 50 pc from the Sun toward TW Hya, Sco-Cen, and Perseus. We believe we will need to approach an angular resolution of nearly 1 AU at the shortest wavelengths toward these systems (20 mas at $\lambda \approx 0.2$ μm). H$\alpha$ and other key spectral line-diagnostics will be used to estimate photo-ionization induced mass-loss rates in irradiated proplyds, giving critical clues to their typical lifetimes.

**Outflows**: Microjets, jets, wide-angle flows, winds, motions, momenta, mass-loss rates, turbulence, shocks.

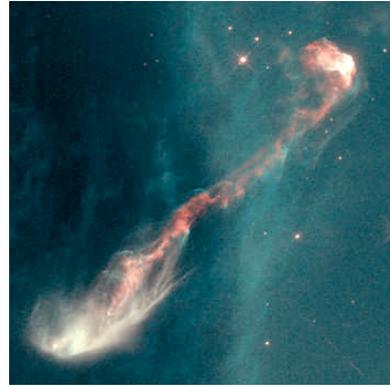

HST has demonstrated that sub-arcsecond imaging is needed to begin to resolve the structure of shocks, and distinguish shock fronts from post-shock cooling layers. Furthermore, only space-based UV/optical observations can measure proper motions on a time-scale short compared to the cooling time. The survey observations will measure the proper motions of hundreds of outflows, enabling the first direct measure of the momentum and energy injected into the ISM by protostellar outflows for a wide-range of stellar masses and star forming environments. Jet orientation changes will trace the history of stellar encounters in clusters. We will also measure the angular momentum of jets to determine their launch points.

While jets and shocks are interesting in their own right, as they emerge from a molecular cloud they also provide a signpost of the youngest protoplanetary accretion disks that are still deeply embedded. The spacing of major ejecta within a single outflow system traces the accretion history of the source YSO. In this way, jet structure provides a fossil record of the accretion and mass-loss histories of the source stars.

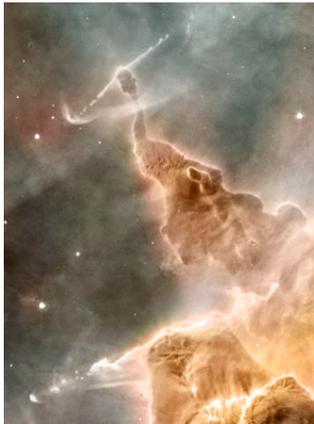

**Nebulae**: Excitation, motion, ionization fronts, triggered star formation;

This imaging program is deliberately designed to investigate the formation of HII regions and expanding bubble systems. How do ionization fronts disrupt surrounding clouds? Under what conditions to they trigger star formation in the medium? High spatial-resolution images with multiple narrowband filters are essential to resolve and correctly model the complex stratified structure of an ionization front. Each HII region / OB association provides a snapshot in time of a range of evolutionary stages. The portions of each region closest to the massive stars are likely to be the most evolved, oldest, and most processed parts of each region. As one moves away from the center, the gas, stars, and disks are likely to be in a younger evolutionary state.

**Massive stars**: Motions, variations, winds, interactions with siblings, HII regions

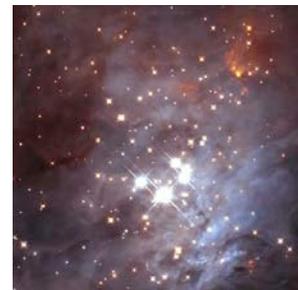

The program will also investigate stellar wind bubbles in HII regions and the interactions of stellar winds with cometary clouds, proplyds, naked young stars and their winds and jets, and the surrounding ISM. Another goal will be to investigate the properties of C-symmetric jets and outflows, wind-jet interactions, supernova-protostellar jet interactions in Orion, Carina, Rosette, NGC 3576, and other regions.

**Recycling**: Supernova remnants and planetary nebulae, bulk motions, excitation, shocks.

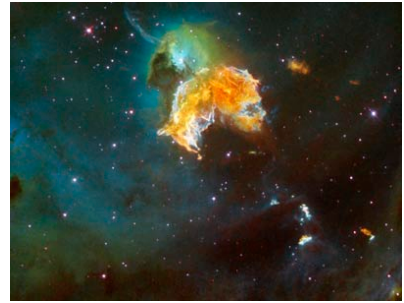

The late stages of stellar evolution - especially in massive stars - are an integral piece of the star and planet formation puzzle, because outflows from the deaths of massive stars drive the chemical evolution and energetics of the ISM. In particular, supernova ejecta enrich the ISM with the elements needed for life to exist, while supernova shocks and stellar winds may compress the surrounding ISM to trigger new star formation. Outflows from the deaths of intermediate mass stars (planetary nebulae) also enrich the ISM with dust, which is vital to the formation of molecular clouds.

By studying the structure and proper motions of a representative sample of nearby supernova remnants (the Crab, IC443, Cas A, Vela, the Cygnus Loop, etc), WR star bubbles (NGC6888, NGC2359), and planetary nebulae (the Helix, M27, etc), the fine details of the shocks and ionization fronts can be spatially resolved. The supernova remnants IC443 and Vela are particularly interesting, as they are directly interacting with molecular clouds. Also, this survey will probe unique regions such as Carina and NGC3603, where the stars are so massive and their lifetimes so short that their imminent death (Eta Carinae and Sher 25) is directly affecting the birth of stars in the same region.

Altogether, these data will probe the disruption of clouds, the recycling of stellar ejecta, and the compression of the ISM into new generation of clouds and the triggering and propagation of star formation.

**Superbubbles**: Destruction of clouds, OB associations, T associations, global structure and evolution of star forming regions.

The energy input from the combined influence of UV radiation, stellar winds, and supernovae from massive stars makes "swiss cheese" out of the ISM. In the most massive star forming regions, where dozens of OB stars live fast and die young before moving very far from their birth sites, the combined effect of this feedback can blow giant shells or "superbubbles" that may eventually break out of the galactic plane, driving a galactic fountain that is vital to the recycling of the ISM. In a few regions such as Carina, NGC3603, NGC3576, W1, and W4, we have the opportunity to study the formation of superbubbles in exquisite detail, where we can actually resolve the structure of the expanding bubbles and model their physical properties.

**The Galactic Ecology**: Impact of spiral arms, formation of clouds, Galactic gradients in YSO and cluster properties, the Galactic Center

We believe an investigation of the "galactic ecology" is vital to understanding the global nature of the star formation process - the formation of giant molecular clouds from the ISM. How do HII regions and superbubble ionization fronts compress the surrounding ISM? Does ram-pressure trigger cloud formation? How do spiral arms trigger cloud formation? How do clouds and cloud cores collapse into clusters, and multiple stars?

**Key Advances in Observation Needed**

To achieve the science goals of this program a variety of capabilities need to be implemented. The majority of the tracers and the various phases of the ISM and stellar populations being targeted require the angular resolution and wavelength agility of a medium to large aperture (1.5-4m) UV/optical space telescope combined with a wide-field imaging camera that can provide diffraction-limited images into the UV-blue to capture the UV-bright stellar populations that HST has been unable to reach. Such a telescope needs to be located in an orbit that is both dynamically and thermally stable (such as L2) to produce the photometric stability required by many aspects of the science goals. A broad complement of both broad- and narrow-band filters will be necessary to isolate and measure not only the unique tracers of specific atomic species but also the trends in stellar color across entire swaths of our local Galactic neighborhood.

Over the next decade the specific technological capabilities that need to be developed include the ability to construct large focal plane arrays that are flight rated for space in a reliable and straightforward fashion that simultaneously mitigates the risk, maximizes the yield rate and keeps the costs down. This is a major challenge that affects not only this project but many others, and requires real investment on the part of the community to allow such systems to be built routinely. In addition, the design of next generation coatings and dichroic optical elements will allow for the design and construction of truly advanced telescope/camera systems that can yield remarkable advances in imaging efficiency for a minimal investment.

**Four Central Questions to be Addressed**

1. What is the formation and survival rate of Solar System class objects in massive star forming regions? There is a growing body of evidence that many stars form in these environments, and that our own Sun was one such system, based on meteoritic evidence concerning $^{60}$Fe.
2. What is the role of triggering and feedback in star formation propagation? A wide range of predictions from numerical simulations describe the role of triggering and feedback as being anything from dominant to negligible. What is the correlation between environment and the nature of the stellar population that forms in secondary and even tertiary star formation events?
3. How is the distribution of star formation across a galactic disk managed? We see evidence that an increase in the efficiency or intensity of star formation occurs almost simultaneously across large distances – what is the source of these global modes – what environmental changes are necessary to initiate and support star formation at these levels?
4. When considering global star formation, what are the determining factors that cause stars to form in one place as opposed to another? At the microphysics level, how does elevated or starburst star formation compare to the more common modes? What dictates the intrinsic efficiency of the star formation process? These latter questions will require comparison with observations from other nearby galaxies such as the LMC, but the database of observations from this program will be necessary to lay the groundwork to answer them.

**Area of Unusual Discovery Potential for the Next Decade**

While the science program in this paper have defined a loose set of specifications (see Table 1), it should be recognized that the opportunity for truly unique discovery is made possible by the **combination** of both a wide angular field of view (tens of arcminutes on a side) **with** the diffraction limit of a medium to large aperture in the UV/optical (resolution elements below 10-20 mas). HST and JWST have provided and will provide exquisite resolution but over very small fields of view. Many problems, such as those discussed here, and others such as the nature of the Universe around the time of Reionization, require not only large collecting area and high resolution, but large fields of view to locate and measure very rare objects, or suites of objects whose location cannot be known a priori. The potential discovery rate from such a combined capability cannot be underestimated, and should be very seriously considered by the Decadal Survey.

| Parameter | Specification | Justification |
|---|---|---|
| Field of View | At least 200 sq. arcmin | To allow a statistically complete survey of as many targets and environments as possible in a reasonable period of time |
| Resolution | Diffraction Limited to 300nm | To provide access to UV-blue stellar populations; to resolve structure in YSO jets, protoplanetary disks, ionization fronts, etc. |
| Aperture | 1.5-4m | This is driven by the limiting surface brightnesses and magnitudes needed traded against the necessary exposure times to achieve them – the larger the better |
| Stability | A small percentage of a pixel | To allow the stable photometry and astrometric measurements necessary to achieve the science goals |
| Photometric Stability | Combination of gain, A/D conversion and QE need to be stable to better than $10^{-5}$ | Again to provide the photometric stability to achieve the science goals of the project |
| Filter Suite | F250W, F336W, F438W, F625W, F775W, F850W; F547M, F980M, F1020M, F1050M, F1080M; F280N, F373N, F469N, F487N, F502N, F631N, F656N, F673N, F953N | Dictated by both broad-band colors needed to survey stellar populations and the narrow-band diagnostics necessary to probe the resolved gas structure and dynamics |
| Optical Design | Efficient design offering a wide, well-corrected field of view to be populated by a large focal plane | The science program can only be achieved by an efficient design that offers parallel observing in the red and blue, with little field distortion, and as large an objective as possible |
| Detectors | High yield, efficient detectors, customized in their response to the passbands needed | Tiling the large focal plane will be challenging – we need an efficient manufacture and testing process, combined with the ability to match response to the optical channels |

**Table 1**: Science Driven General Specifications